# A Digital Twin-Based Method for Evaluating Local Collective Tariffs in Distribution-Level Energy Systems

Christensen, Kristoffer; Jørgensen, Bo Nørregaard; Ma, Zheng Grace

# A Digital Twin-Based Method for Evaluating Local Collective Tariffs in Distribution-Level Energy Systems

Kristoffer Christensen[1*[0000-0003-2417-338X]], Bo Nørregaard Jørgensen[1[0000-0001-5678-6602]] and Zheng Grace Ma[1*[0000-0002-9134-1032]]

[1] SDU Center for Energy Informatics, Maersk Mc-Kinney Moeller Institute, The Faculty of Engineering, University of Southern Denmark, Odense, Denmark

kric@mmmi.sdu.dk, bnj@mmmi.sdu.dk and zma@mmmi.sdu.dk

**Abstract.** This work addresses the need for engineering-grounded evaluation of implemented tariff mechanisms in distribution-level energy systems. A digital twin-based method is proposed for assessing local collective tariffs under realistic behavioral and infrastructural conditions. The approach integrates agent-based modeling of household consumption and generation, virtual aggregation through a shared metering abstraction, and explicit representation of tariff logic within a unified simulation environment. The method is demonstrated through its application to the Danish Local Collective Tariff across representative residential energy community configurations, including scenarios with photovoltaic generation, battery storage, and electric vehicle charging. Results indicate that aggregation of heterogeneous demand profiles reduces peak coincidence and enables more efficient allocation of tariff components, leading to measurable cost reductions at the community level. At the same time, the outcomes reveal sensitivity to the temporal alignment of consumption and generation, influencing the degree of cost neutrality across participants. The findings illustrate how digital twin-based evaluation can support systematic assessment of tariff mechanisms by capturing the interaction between infrastructure, user behavior, and regulatory design. The proposed approach provides a basis for analyzing and comparing tariff structures in distribution-level energy systems beyond the specific case considered.

**Keywords:** digital twin, agent-based simulation, energy communities, tariff design, distribution-level energy systems, local collective tariffs, energy informatics, demand aggregation

## 1 Introduction

This paper addresses an emerging question in electricity distribution tariff design, namely how a localized tariff should be evaluated when it is introduced into a real and already regulated market setting. From January 1, 2026, the Danish distribution system operators Cerius and Radius introduced a Local Collective Tariff (LCT) for eligible low-voltage customer groups connected behind the same 10/0.4 kV transformer [1-3].

The method is intended to create incentives for better simultaneity between local consumption and local production while preserving cost reflectivity and leaving ordinary customers outside the scheme on Tariff Model 3.0 [1, 4].

From a research perspective, the policy is significant because it represents an implemented localized tariff in a real regulatory setting, introduces virtual collective measurement and settlement, and links tariff design with local energy sharing and business ecosystem coordination.

The present study extends two previously developed models. Christensen et al. developed a multi-agent simulation of electric vehicle adoption and its impacts on a Danish low-voltage distribution grid with 126 residential consumers [5]. Værbak et al. developed a digital twin framework for simulating distributed energy resources in distribution grids and demonstrated how communication, grid representation, and consumer models can be integrated in a modular simulation environment [6]. Building on these foundations, the present work introduces the tariff logic of a LCT, adds a virtual meter representation in Energinet's DataHub, and models an Energy Sharing Consultant as a new ecosystem actor inspired by the business ecosystem design method of Ma et al. [7].

The case study uses Strib as a representative Danish residential transformer area with 126 consumers connected to a 400 kVA transformer, real household demand data from 2019, and weather data from a nearby station. The purpose is not to estimate power flow or voltage quality effects in the present paper. Rather, the immediate objective is to compare annual and monthly economic outcomes under Tariff Model 3.0 and the new LCT under a set of carefully defined scenarios. This economic framing is motivated directly by the LCT method documents, which state that the tariff is calibrated to ensure that, without behavioral changes, local collectives pay the same revenue as if the customers were billed individually under Tariff Model 3.0 [2, 4].

Existing approaches to tariff evaluation typically rely on optimization-based formulations, stylized market simulations, or purely analytical models. These approaches often assume idealized behavior, lack representation of distribution-level physical constraints, or abstract away from the operational realities of implemented tariff schemes. As a result, a gap emerges between regulatory design and operational validation. There is a need for engineering-oriented evaluation methods capable of representing the interaction between infrastructure, user behavior, and tariff logic within realistic system configurations.

This work addresses the lack of engineering-grounded evaluation methods for implemented tariff mechanisms in distribution-level energy systems. The primary contribution is a digital twin-based methodological framework for evaluating LCT structures under realistic behavioral and infrastructural conditions. The framework integrates physical grid constraints, agent-level consumption and production behavior, and regulatory tariff logic within a unified simulation environment. A secondary contribution is the empirical evaluation of the Danish LCT using this framework across representative residential energy community configurations. A third contribution is the identification of system-level mechanisms that govern cost allocation, neutrality, and sensitivity under shared tariff structures, providing insight into how aggregation and coincidence effects influence economic outcomes.

The remainder of the paper is structured as follows. Section 2 reviews related work. Section 3 presents the research design. Section 4 explains the case study, tariff representation, and scenario design. Section 5 presents and discusses the results. Section 6 concludes the paper.

## 2 Related Work

The related work was screened with the aim of identifying studies that are most relevant for evaluating implemented localized tariff mechanisms in distribution-level energy systems. Priority was therefore given to studies addressing: (i) distribution tariff design and local grid cost allocation, (ii) local energy communities and energy sharing arrangements, and (iii) digital twin or simulation-based approaches for distributed energy resources in low-voltage systems. Studies focusing only on wholesale markets, generic electricity pricing, or high-level policy discussion were excluded unless they provided specific insight into tariff structure, settlement logic, or distribution-grid implications. This screening supports a focused comparison between existing tariff and local energy community studies and the present digital twin-based evaluation of an approved Danish Local Collective Tariff.

### 2.1 Localized tariffs, local energy communities, and distribution tariff design

Localized tariffs and local electricity market arrangements have attracted growing attention because conventional volumetric tariffs often fail to represent the network implications of distributed energy resources, flexible demand, and prosumer coordination [8-11]. The literature now distinguishes between at least three related but not identical domains: local energy communities that aim to maximize local self-consumption, peer-to-peer or local energy markets that allow intra community transactions, and distribution tariff reforms that seek to recover network costs while giving more accurate locational or capacity related signals [8, 9, 11].

Maldet et al. review trends in local electricity market design and show that grid tariffs remain a central regulatory issue because local exchange can reduce billed imports without necessarily reducing network cost drivers [8]. Their review is especially relevant to the present paper because it stresses that tariff architecture cannot be treated as a secondary detail. Rather, the economic value of local exchange depends critically on how network use is measured and charged. This observation aligns closely with the Danish LCT, where the incentive structure emerges from the combination of lower time differentiated energy tariffs and an explicit power related charge [1, 4].

Rahman et al. review customer centric dynamic tariff design and emphasize that tariff signals must balance efficiency, comprehensibility, fairness, and implementation feasibility [10]. While their review is broad and not limited to local collectives, it is useful here because LCT combines several of these design dimensions. It introduces a stronger capacity related signal, retains time dependent volumetric elements, and must remain administratively feasible for residential users. However, these studies primarily evaluate tariff mechanisms under simplified assumptions regarding user behavior and system configuration, and do not provide engineering-level validation of implemented tariffs within realistic distribution network contexts.

Azim et al. further show that tariffs, operational management, and market design for distributed energy resources are increasingly interdependent, especially when distributed generation, electric mobility, and local flexibility become widespread [11]. This broader review helps situate the current study within a larger transition from passive retail tariffs toward tariffs that coordinate behavior, investment, and operational flexibility.

## 2.2 Evidence from energy communities and local market studies

Several studies analyze tariff and pricing effects inside local communities, but most do so in stylized optimization settings rather than in implemented tariff regimes. Lüth et al. study distributional effects in local electricity market designs and show that gains can be unevenly distributed among participants unless tariff and market rules are carefully chosen [12]. Their work is important because it highlights that collective welfare improvements can coexist with problematic internal allocation outcomes. The present study inherits this concern and therefore models internal settlement rules explicitly, even though the primary comparison is made at group level.

Manso Burgos et al. examine Spanish local energy communities under the country's new tariff structure and variable sharing coefficients [13]. They show that tariff details materially influence economic outcomes and that community performance depends on the fit between production patterns, demand timing, and regulatory allocation rules. The present study differs in two key ways. First, it addresses a Danish tariff that is explicitly designed around the same transformer station and a collective virtual measurement point. Second, it evaluates a tariff already approved for real implementation rather than a generalized regulatory option.

Velkovski et al. analyze how different tariff structures affect energy community operation and grid related indicators using a mixed integer optimization framework [14]. Their results demonstrate that tariff choice changes both internal scheduling and network relevant outcomes. Yet their setting remains optimization based and assumes active operational response. By contrast, the current paper first evaluates the baseline claim of cost neutrality under unchanged behavior, which is a precondition for interpreting later flexibility or control responses.

Lersch et al. investigate energy sharing in low voltage distribution grids in Germany and demonstrate that energy sharing may improve some local objectives while also affecting voltages, loading, and reinforcement needs under future electrification [15]. Their paper is particularly relevant because it connects energy sharing to physical grid performance rather than only participant cost. However, it does not focus on the Danish LCT method or on the specific business and settlement architecture introduced by a virtual meter and consultant actor. The present work therefore complements such studies by focusing first on tariff realism and institutional implementability below the transformer level.

While these works analyze economic and operational aspects of energy communities, they do not explicitly address how implemented tariff structures interact with aggregated consumption and production profiles under regulatory constraints.

### 2.3 Modeling approaches and digital twin perspectives

Digital twins and simulation frameworks are increasingly used to study distribution grids with distributed energy resources because they can represent interacting technical, economic, and behavioral processes [6, 16]. Vecchi et al. review modeling tools for renewable energy communities and show a strong need for approaches that combine technical realism with tractable representation of regulatory and market rules [16]. This is directly aligned with the current study, where tariff design is inseparable from technical aggregation at transformer level and from settlement logic in the market layer.

Within the authors' prior work, Christensen et al. demonstrate the value of multi-agent-based simulation for representing household heterogeneity and electric vehicle adoption in a 126 household Danish distribution grid [5]. Værbak et al. extend the modeling perspective by proposing a digital twin framework with modules for distributed energy resources, consumers, the grid, and energy management [6]. These works form the technical basis for the present study. However, neither paper models the newly introduced LCT nor the associated market roles and virtual meter concept.

Ma et al. propose a methodology for business ecosystem architecture development and argue that actor roles, interactions, and value exchanges should be made explicit when complex energy services are introduced [7]. This perspective is highly relevant to LCT, because tariffing is no longer merely a bilateral relation between the DSO and an individual household. Instead, it involves the DSO, Energinet's DataHub, a collective legal entity, ordinary suppliers, and a coordinating actor that can administer data flows and billing. Modeling this ecosystem explicitly is one of the distinctive contributions of the present work.

In this context, the digital twin is not limited to a technical replica of the low-voltage grid. Rather, it represents the distribution-level energy business ecosystem associated with the LCT, including the relevant actors, roles, objects, and interaction flows. These include monetary, goods and service, data, and information flows required to evaluate the LCT as an implemented socio-technical arrangement.

### 2.4 Research gap and contribution

The literature reviewed above establishes that local tariff structures matter, that internal allocation is nontrivial, and that digital twin style models are useful for studying distributed energy resources. Nevertheless, a clear gap remains. Existing studies usually analyze hypothetical local markets, optimization-based coordination schemes, or generalized regulatory options [8, 12-15]. Much less attention has been given to the engineering evaluation of a newly implemented and regulator approved localized tariff using a case specific digital representation of the actual actor ecosystem below a real low-voltage transformer.

This paper addresses that gap. Its first contribution is methodological: it extends an existing multi-agent and digital twin framework with a virtual meter and Energy Sharing Consultant to represent tariff logic and ecosystem interactions alongside household demand, PVs, and electric vehicles [5-7]. Its second contribution is empirical: it evaluates a real tariff reform introduced in Denmark in 2026 using an approved method architecture [1-4]. Its third contribution is analytical: it isolates cost neutrality under unchanged behavior and examines how aggregation affects economic outcomes.

# 3 Research Design

## 3.1 Research questions and hypotheses

The study is structured around two research questions.

RQ1: Does the LCT produce approximately the same annual network related cost as Tariff Model 3.0 when the underlying hourly demand and production behavior are unchanged?

RQ2: How sensitive is the economic outcome to the number of participating members and to the composition of distributed energy resources, especially Photovoltaic (PV) systems and electric vehicles?

Hypothesis H1 states that, when hourly consumption and production profiles are unchanged, the annual group cost under LCT should remain close to the Tariff Model 3.0 outcome because the method is explicitly designed for this purpose [2, 4].

Hypothesis H2 states that the economic attractiveness of LCT will vary with membership composition because the virtual meter aggregates coincidence effects. In particular, more diverse consumption profiles and higher local PV coincidence may reduce imported energy volumes and collective peaks, but the rolling historical peak basis will delay full realization of any benefit.

The hypotheses are grounded in the expectation that aggregation reduces the temporal coincidence of peak demand across households, thereby lowering peak-based tariff components. Under conditions of diverse consumption profiles and distributed generation, the aggregated load profile is expected to exhibit reduced variance relative to individual profiles, which may lead to more efficient tariff allocation and improved cost neutrality.

## 3.2 Case study and scope

The case study is the residential area of Strib, previously used in [5], with 126 household consumers connected to a 400 kVA 10/0.4 kV transformer, consistent with the aggregation level of the LCT.

Household demand is based on measured 2019 data. Seven consumers are identified as having PV systems due to periods of negative net load. Additional PV systems are assigned in selected scenarios using existing simulation inputs, with installed capacities sampled between 3 and 6 kW in line with typical Danish household systems [17]. Weather data from 2019 is used to generate production profiles.

The analysis is economic rather than electrical. The grid is represented using a simplified direct current model with unit power factor, without explicit modeling of voltage, reactive power, or phase imbalance.

Although Strib is located in DK1, tariff and spot price assumptions are based on Cerius and DK2, reflecting the implemented LCT. The case study is therefore used as a representative dataset for evaluating tariff behavior rather than reconstructing historical billing.

### 3.3 Model extension: virtual meter and Energy Sharing Consultant

The simulation is extended with two elements required to represent the LCT as an operational arrangement. A virtual meter aggregates hourly imports and exports across group members while individual meters remain for taxes and supplier settlement [1, 2]. The virtual meter is conceptually located in Energinet's DataHub.

The virtual meter can be interpreted as the implementation of an aggregation operator, which combines individual household measurements into a collective representation used for tariff calculation. This abstraction enables the simulation of regulatory mechanisms that operate on shared consumption and production profiles without requiring physical aggregation infrastructure.

In addition, an Energy Sharing Consultant is introduced as a model actor responsible for managing information and settlement within the collective. Based on aggregated and individual consumption and production data, the consultant computes monthly member level costs according to the defined internal allocation rules.

These extensions enable representation of both the technical aggregation and the associated business ecosystem underlying the LCT. More generally, the approach constitutes a digital twin-based evaluation framework in which individual household states are aggregated through a virtual meter and mapped to tariff outcomes, enabling consistent analysis of how tariff structures interact with consumption patterns, local generation, and aggregation effects.

## 4 Tariff Representation and Scenario Design

The virtual meter is defined such that, for each hour, import corresponds to the positive difference between total member consumption and total member production, while export corresponds to the positive difference between total member production and total member consumption. This formulation follows the approved method, where net delivered and net injected energy are registered at the collective level on an hourly basis [1]. An important implication is that internally balanced energy is not registered as an exchange with the grid.

The counterfactual benchmark is Tariff Model 3.0 for Cerius in 2026, while the alternative is the corresponding LCT. Both tariff structures are implemented directly from the official 2026 price list, excluding VAT [3]. The tariff parameters used in the simulation are summarized in Table 1.

**Table 1.** Summary of tariff parameters used in the simulation for Cerius 2026, excluding VAT [3].

| Tariff model | Energy rate from 00:00 to 06:00 [Ore/kWh] | Energy rate from 06:00 to 17:00 and 21:00 to 00:00 [Ore/kWh] | Energy rate from 17:00 to 21:00 [Ore/kWh] | Power payment [DKK/kW/year] | Feed-in tariff [Ore/kWh] | Subscription fee [DKK/year] |
|---|---|---|---|---|---|---|
| Tariff Model 3.0 Winter | 10.65 | 31.94 | 95.82 | Not applicable | 1.41 | Consumer: 577.98 Prosumer: 598.43 |
| Tariff Model 3.0 Summer | 11.53 | 17.30 | 44.98 | Not applicable | 0.84 | Consumer: 578.95 Prosumer: 591.33 |
| Local Collective Tariff Winter | 3.79 | 11.37 | 34.10 | 590.05 | 1.41 | 2375 |
| Local Collective Tariff Summer | 4.43 | 6.64 | 17.27 | 680.57 | 0.84 | 2375 |

The essential structural difference between Tariff Model 3.0 and the LCT lies in the introduction of a collective power-based component and reduced time-differentiated energy rates for the virtual meter. While Tariff Model 3.0 relies solely on time-dependent energy charges and subscriptions at household level, the LCT shifts part of the cost structure toward a shared capacity-based signal [1, 2, 4]. The approved method retains the waterfall principle from Tariff Model 3.0 and maintains the local collective as a subset of C customers [1].

Within this structure, the collective power component represents approximately 75% of the relevant variable tariff revenue, while the remaining 25% is assigned to the reduced energy component, excluding losses that remain in the energy element [1]. This design shifts the primary economic signal toward collective peak reduction while preserving temporal differentiation in energy pricing.

The power-based component is determined using a rolling peak mechanism, defined as the average of the ten highest hourly consumption values over the previous twelve months [4]. In the simulation, this average is calculated based on the available historical data when a full twelve-month period is not present. As illustrated in [2], this mechanism implies that peak events can influence the tariff for several subsequent months.

Consequently, reductions in peak demand are not immediately reflected in the cost, as historical peaks remain part of the rolling basis.

The distinction between official tariff rules and author-defined assumptions is important. The virtual collective metering principle, reduced LCT energy tariffs, feed-in tariffs, subscription fee, and rolling peak-based power payment are implemented from the official method and price documents [1-4]. In contrast, the internal member-level settlement rule is an author-defined assumption, since the detailed internal allocation between participants is not prescribed. In the model, internally balanced energy is allocated according to hourly consumption, while the collective power charge is shared equally among LCT members. Excess production is settled internally at the hourly day-ahead spot price; when the spot price is negative, the internal settlement value is set to zero. Remaining excess production exported to the grid is allocated to producers in proportion to their production share and compensated using the spot price and applicable feed-in tariff. The rolling peak charge is initialized using the available simulated history when a full twelve-month window is not yet available, which makes the first months sensitive to the initialization of the rolling peak basis.

Based on this formulation, eight scenarios are constructed to evaluate the economic effects of the LCT under different participation levels, PV penetration rates, and electric vehicle adoption. All scenarios are based on the same 126-household transformer context unless otherwise specified, and are summarized in Table 2.

**Table 2.** Scenario design used in the simulation study, where seven PVs are minimum due the used dataset.

| Scenario | LCT members | PV households | EV households | Tariff regime | Purpose |
|---|---|---|---|---|---|
| S1 | 0 (0%) | 7 | 0 | Tariff Model 3.0 | Baseline without LCT |
| S2 | 126 (100%) | 7 | 0 | All members under LCT | Test full participation with existing PV count |
| S3 | 63 (50%) | 7, all included in LCT | 0 | Partial LCT | Test 50% participation while keeping all PV members inside the collective |
| S4 | 31 (25%) | 7, all included in LCT | 0 | Partial LCT | Test 25% participation with all PV members included |
| S5 | 10 (8%) | 7, all included in LCT | 0 | Small LCT group | Test a very small collective |
| S6 | 126 (100%) | 31 (25%) | 0 | All members under LCT | Test higher PV penetration, about |
| | | | | | 25% of households |
| S7 | 0 (0%) | 7 | 23 (18%) | Tariff Model 3.0 | Test EV load impact without LCT |
| S8 | 126 (100%) | 7 | 23 (18%) | All members under LCT | Test EV load impact with LCT |

The scenario logic is cumulative and comparative. S1 and S2 provide the main neutrality test. S3 to S5 test sensitivity to membership size while preserving local generation inside the collective. S6 tests whether broader PV penetration changes the tariff outcome when all households participate. S7 and S8 test whether a substantial electrification load represented by 23 electric vehicles, corresponding to 18% of households, changes the relative attractiveness of LCT.

The EV share of 18% is used as a representative scenario parameter based on Danish fleet statistics [18, 19].

EV charging follows a simple charge-at-arrival logic. The additional PV systems in S6 are a scenario input used to stress the tariff mechanism under stronger local production–demand coincidence. Because batteries are excluded, the comparison isolates direct simultaneity and group aggregation.

# 5 Results and Discussion

This section evaluates the economic implications of the LCT across the eight simulation scenarios. The analysis focuses on aggregated electricity expenses, aggregated income from surplus production, and, where applicable, the cost allocated to the LCT group. The analysis is structured around three themes: hypothesis validation, group composition, and distributed energy resources.

## 5.1 Overview of Scenario Outcomes

Table 3 summarizes the main economic results for all scenarios. The total electricity expenses represent the aggregated annual cost for the full case study area. The income values represent the aggregated annual revenue from production exported to the grid or settled externally. For scenarios with an active LCT, the final four columns show the aggregated cost metrics for the participating LCT group only.

**Table 3.** Economic summary of the eight scenarios. Negative values represent costs.

| Scenario | Description | Total electricity expenses, sum [DKK] | Mean user expense [DKK] | Minimum user value [DKK] | Income, sum [DKK] | Mean income per prosumer [DKK] | LCT group cost, sum [DKK] | LCT group mean [DKK] |
|---|---|---|---|---|---|---|---|---|
| S1 | Baseline, 7 PV, no LCT | -492,752.0 | -3,910.7 | -10,034.3 | 5,767.5 | 823.9 | N/A | N/A |
| S2 | 126 users in LCT, 7 PV | -320,202.0 | -2,541.3 | -7,091.7 | 5,696.1 | 813.7 | -320,202.0 | -2,541.3 |
| S3 | 63 users in LCT, all 7 PV included | -409,451.4 | -3,241.7 | -9,083.6 | 5,691.9 | 813.1 | -150,571.0 | -2,390.0 |
| S4 | 31 users in LCT, all 7 PV included | -448,018.8 | -3,555.7 | -9,083.6 | 5,756.4 | 822.3 | -81,606.1 | -2,632.5 |
| S5 | 10 users in LCT, all 7 PV included | -478,156.2 | -3,794.9 | -9,083.6 | 5,623.2 | 803.3 | -23,236.8 | -2,323.7 |
| S6 | 126 users in LCT, 25% PV penetration | -277,949.8 | -2,206.0 | -7,104.5 | 30,344.0 | 978.8 | -277,949.8 | -2,206.0 |
| S7 | 126 users, 7 PV, 23 EV, no LCT | -635,560.3 | -5,044.1 | -17,290.3 | 5,742.5 | 820.4 | N/A | N/A |
| S8 | 126 users, 7 PV, 23 EV, with LCT | -424,340.2 | -3,367.8 | -12,411.8 | 5,667.3 | 809.6 | -424,340.2 | -3,367.8 |

A first observation is that all scenarios with full LCT participation show lower aggregated electricity expenses than their corresponding non-LCT counterparts. This applies both to the base PV configuration and to the electric vehicle configuration.

This reduction can be attributed to decreased coincidence of peak demand within the aggregated load profile. By combining heterogeneous consumption patterns, the community-level peak becomes lower than the sum of individual peaks, resulting in reduced exposure to peak-based tariff components.

The higher mean cost in S4 is explained by the presence of a high consumption user in the LCT group. This user has an annual consumption of approximately 8,200 kWh, compared to a mean of approximately 3,200 kWh across all consumers. As a result, the aggregated peak demand of the group increases, which raises the power-based tariff component. Since this cost is shared equally among group members, the mean group expense becomes higher than in both S3 and S5. In S3, the effect is distributed across more members, while in S5 this user is not part of the LCT group.

### 5.2 Hypothesis Assessment and Scenario Effects

The central hypothesis H1 of the study is that the LCT should lead to the same overall cost as Tariff Model 3.0 if consumption behavior is unchanged. This is assessed by comparing the baseline case without LCT in S1 and the full participation LCT case in S2, where the underlying consumption behavior is identical.

The comparison shows a substantial difference in aggregated electricity expenses. Scenario S1 yields total expenses of -492.8k DKK, while S2 yields -320.2k DKK. This corresponds to an improvement of approximately 172.6k DKK at the system level. The mean user expense is likewise reduced from -3,911 DKK in S1 to -2,541 DKK in S2.

Under the present implementation assumptions, hypothesis H1 is therefore not confirmed. The LCT does not reproduce the same annual cost as Tariff Model 3.0 under unchanged behavior. Instead, the model indicates a clear economic advantage for the LCT group.

The observed deviations from neutrality reflect the sensitivity of cost allocation to the temporal alignment between generation and consumption. When local generation coincides with high demand periods, the benefits of aggregation are amplified, whereas misalignment reduces the effectiveness of shared tariff structures.

This difference reflects the combined effects of internal production settlement, shared power-based charges, and value transfers associated with spot-price-based internal settlement.

These results also provide support for hypothesis H2. The economic outcome of the LCT is sensitive to both participation level and group composition. As shown in scenarios S3 to S5, reduced participation leads to higher total system costs due to less effective local balancing, while the mean cost within the LCT group depends strongly on which users are included. In particular, the presence of a high-consumption user in S4 increases the aggregated peak demand and thereby the shared power-based cost, whereas the absence of this user in S5 results in a lower mean group cost. Furthermore, scenarios with higher PV penetration, such as S6, demonstrate enhanced economic performance due to increased local utilization of production.

Even without behavioral adaptation, tariff design and settlement assumptions alone produce different cost outcomes and remain strongly dependent on group composition and participation structure.

Fig. 1 shows the load and rolling peak values in the virtual meter for scenarios 2 and 3. The zero peak in the first month results from the rolling-window calculation, which requires prior observations. Negative hourly values in scenario 3 indicate periods of surplus production, which are exported and settled at spot price plus feed-in tariff in proportion to each producer’s share.

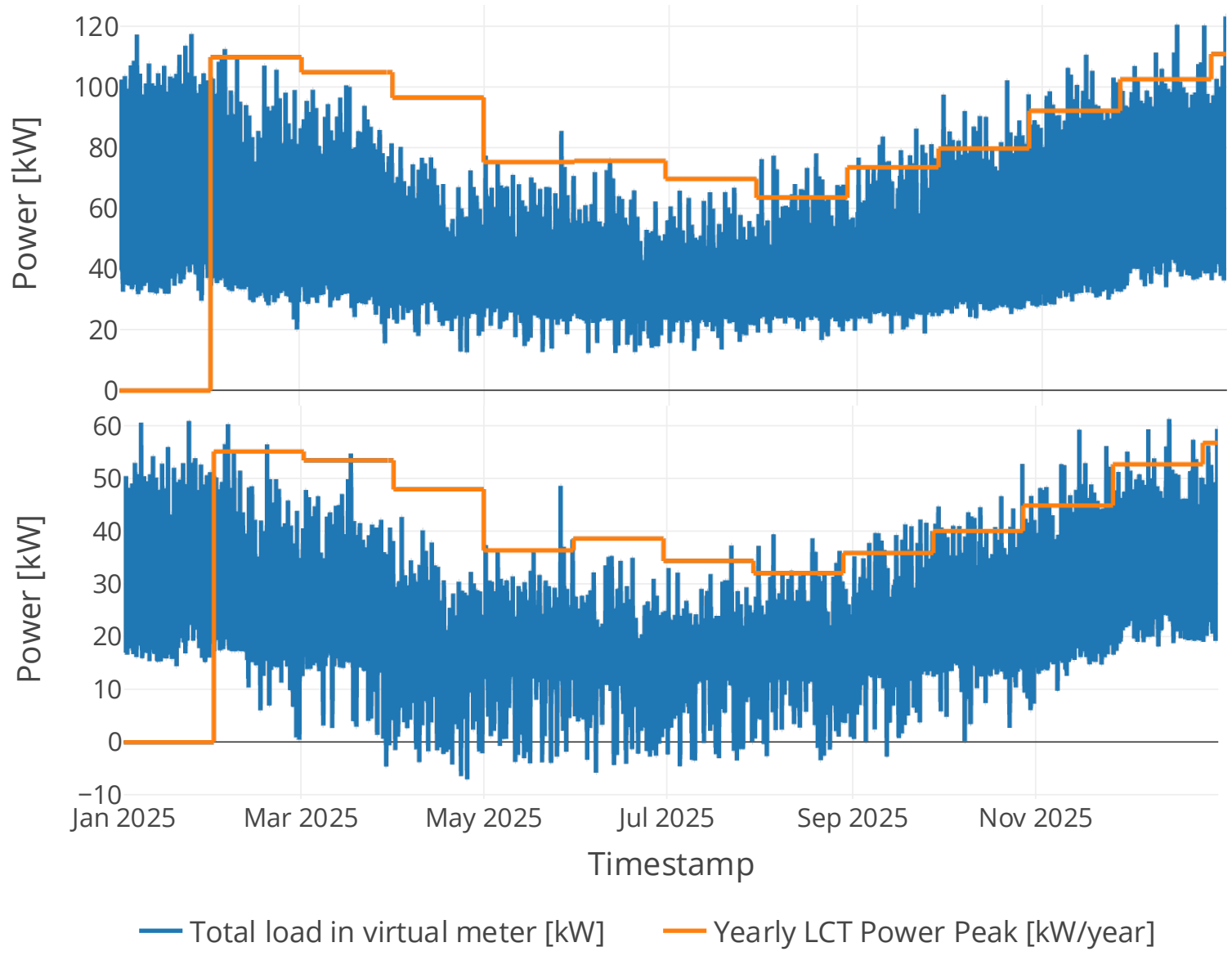

**Fig. 1.** Total load and calculated power peak in virtual meter for scenario 2 with 100% LCT (top) and for scenario 3 with 50% LCT (bottom).

Scenarios S3, S4, and S5 investigate the effect of reduced participation in the LCT while keeping all seven PV users inside the LCT group. The total system expenses increase as the LCT group size decreases. Specifically, total expenses change from -409,451 DKK in S3 to -448,019 DKK in S4 and -478,156 DKK in S5.

Scenario S6 considers full LCT participation with approximately 25% PV penetration. This scenario yields the lowest aggregated electricity expenses among all scenarios, at -277.9k DKK, and the highest aggregated income, at 30,344 DKK. The mean user expense is -2,206 DKK, which is the lowest among the eight scenarios.

The large increase in income confirms that higher local PV penetration substantially increases the value of local energy sharing under the assumed settlement structure. More local production becomes available for internal balancing, which reduces dependence on grid imports and improves local value retention. The result also indicates that the LCT framework becomes more economically attractive when the local energy community includes a larger distributed generation base.

Scenarios S7 and S8 examine the effect of electric vehicle integration, with 23 electric vehicles corresponding to approximately 18% penetration. As expected, the addition of electric vehicles increases annual electricity expenses because of the extra charging load. Fig. 2 shows the load and calculated peak power in the virtual meter in scenario 8, which clearly shows the impact on the peak power compared to scenario 2.

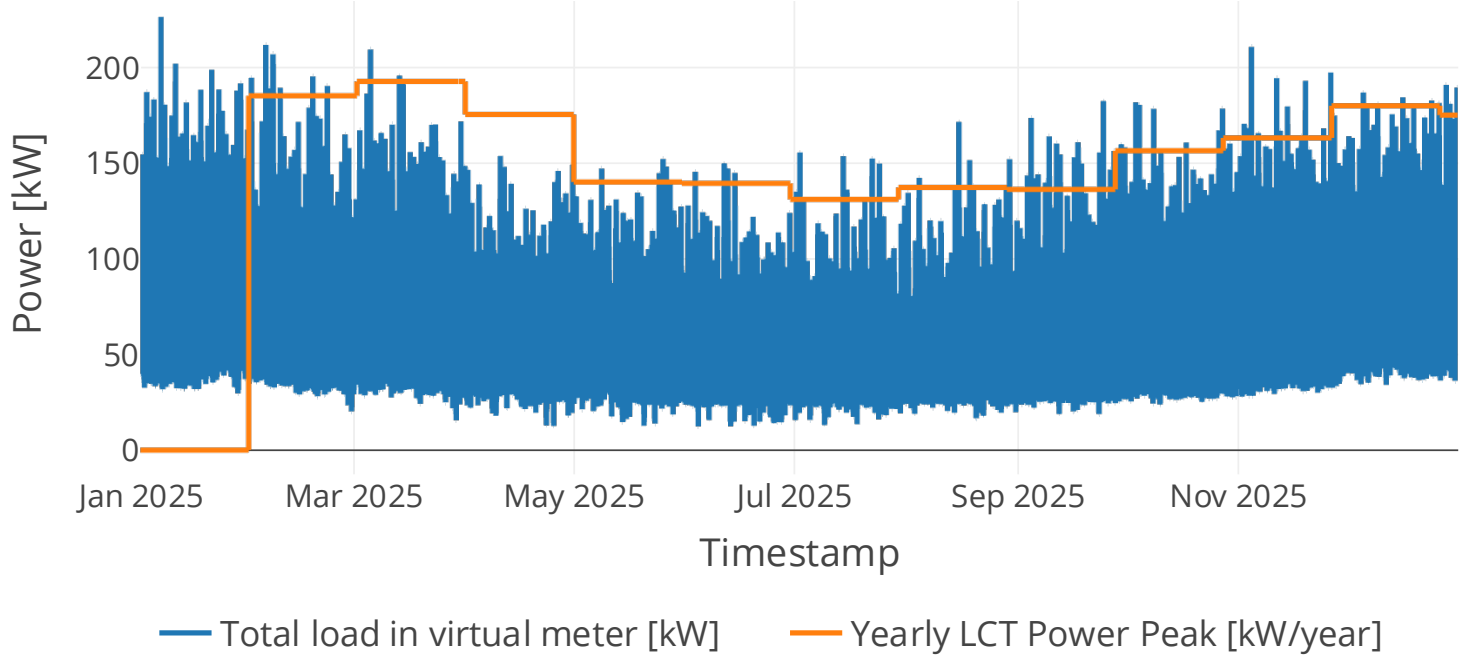

**Fig. 2.** Total load and calculated power peak in virtual meter for scenario 8 with 18% EVs.

Under Tariff Model 3.0 without LCT, S7 produces the highest total expenses of all scenarios, at -635.6k DKK, with a mean user expense of -5,044.1 DKK per user.

When the same scenario is evaluated with full LCT participation in S8, total expenses are reduced to -424.3k DKK, and the mean user expense is reduced to -3,368 DKK. The cost reduction is therefore substantial, both in aggregate and per user terms.

This indicates that the LCT mitigates part of the economic burden introduced by electric vehicle charging. Since the tariff contains a power-based component calculated from aggregated peak demand, the result suggests that collective settlement can smooth the economic effect of higher loads when users are combined under the same virtual meter. Even without active charging control, the LCT appears to provide a more favorable economic framework than the ordinary tariff.

This suggests that collective tariff settlement may support residential electrification, although the present analysis does not include smart charging or centralized control strategies.

### 5.3 Summary of Key Findings

Overall, the LCT changes economic outcomes across all tested configurations. The electric vehicle scenarios further show that the LCT can partially offset the increased cost associated with electrification.

At the same time, the fact that S1 and S2 do not converge economically under unchanged behavior is noteworthy. This suggests that the practical implementation of the tariff and internal settlement mechanism requires further scrutiny. The intended neutrality relative to Tariff Model 3.0 may depend on assumptions that are difficult to reproduce in practice, or on settlement conventions that are not yet fully specified in the available method documents.

The difference between S1 and S2 should therefore be interpreted as the combined effect of the official LCT structure and the assumed internal settlement rule, rather than as a pure tariff-price effect. In particular, the equal sharing of the collective power charge can redistribute costs between low- and high-consumption members, while PV ownership and temporal production-consumption alignment affect how benefits from internally balanced energy are allocated. This raises fairness considerations that are only partially captured by the reported minimum and maximum values. A full sensitivity analysis of alternative settlement rules, including proportional power-charge allocation or alternative treatment of negative prices, is outside the scope of the present paper because it would require additional scenario simulations. Similarly, additional distributional indicators such as quartiles, standard deviation, and percentile ranges should be included in future work to assess member-level fairness more systematically. The scenarios also exclude batteries, smart charging, and behavioral adaptation in order to isolate aggregation and simultaneity effects under unchanged behavior. Finally, because Strib demand data are combined with Cerius/DK2 tariff and price assumptions, the case should be interpreted as a representative Danish low-voltage evaluation case rather than a reconstruction of historical billing for Strib.

## 6 Conclusion

This paper evaluated the economic implications of the Local Collective Tariff introduced in Denmark in 2026 using a multi-agent and digital twin framework for a low-voltage residential system with 126 consumers.

The results provide clear answers to the research questions. For RQ1, the Local Collective Tariff does not result in approximately the same annual cost as Tariff Model 3.0 under unchanged consumption behavior. Instead, it leads to a systematic reduction in both total system costs and average user costs under the modeled assumptions.

For RQ2, the economic outcome is highly sensitive to participation level and group composition. Larger groups benefit from increased diversity and improved internal balancing, while the presence of high consumption users can significantly influence cost allocation through the shared power component. Increased PV penetration enhances economic performance, whereas electric vehicles increase demand but exhibit a reduced cost impact under the Local Collective Tariff compared to Tariff Model 3.0.

These findings indicate that the Local Collective Tariff introduces structural changes in cost allocation. The tariff consistently reduces overall costs, with the strongest effects observed for high participation and increased local generation. At the same time, performance depends strongly on group composition, as individual high consumption users can significantly influence aggregated peak demand and shared costs.

From a methodological perspective, the study demonstrates that digital twin-based simulation is suitable for evaluating implemented tariff designs in realistic settings, enabling analysis beyond stylized tariff comparisons by integrating demand data, distributed energy resources, and market roles.

The presented framework enables systematic evaluation of implemented tariff mechanisms within realistic energy community configurations, bridging the gap between regulatory design and operational performance. The findings indicate that local collective tariff structures can improve cost efficiency under conditions of aggregated demand and distributed generation, while remaining sensitive to system configuration and behavioral diversity. The approach provides a basis for further investigation of tariff design, including the integration of flexibility mechanisms, dynamic pricing, and control strategies within distribution-level energy systems.

**Acknowledgement**
This work is partly supported by Nordic Energy Research through the funding of Energy Informatics Academy Network Nordic (EIA Nordic), Project number 198639; Part of the project titled "Danish participation in IEA ES Task 46", funded by EUDP (project number: 134252-556333); Part of EnergyBuilder, grant 134251-549133 funded by Innovation Fund Denmark.

## Declarations

**Conflict of Interest/Competing Interests:**
The authors declare no competing interests

**Ethics approval/Consent:**
Not applicable

**Data availability:**
Not applicable